\documentclass[%
 reprint,
 amsmath,amssymb,
 aps,
showkeys,
]{revtex4-2}

\newcounter{inlineenum}
\renewcommand{\theinlineenum}{\roman{inlineenum}}
\newenvironment{inlineenum}
{\unskip\ignorespaces\setcounter{inlineenum}{0}%
	\renewcommand{\item}{\refstepcounter{inlineenum}{\textit{\theinlineenum})~}}}
{\ignorespacesafterend}

\usepackage{graphicx}
\usepackage{dcolumn}
\usepackage{bm}

\begin{document}

\preprint{APS/123-QED}

\title{Inferring the stability of concentrated emulsions \\
from droplet configuration information}

\author{Danny Raj Masila}
\email{dannym@iisc.ac.in}
\homepage{http://www.dannyraj.com}
\affiliation{%
 Department of Chemical Engineering, IISc Bangalore, Karnataka, India
}%
\author{Pavithra Sivakumar}%
\affiliation{%
 Department of Electrical Engineering, IIT Madras, Tamilnadu, India
}%

\author{Arshed Nabeel}
\affiliation{
 Center for Ecological Sciences, IISc Bangalore, Karnataka, India
}

\date{\today}

\begin{abstract}
When droplets are tightly packed in a 2D microchannel, coalescence of a pair of droplets can trigger an avalanche of coalescence events that propagate through the entire emulsion. This propagation is found to be stochastic, \textit{i.e.} every coalescence event does not necessarily trigger another. To study how the local probabilistic propagation affects the dynamics of the avalanche, as a whole, a stochastic agent based model is used. Taking as input, 
\begin{inlineenum}
    \item how the droplets are packed (configuration) and
    \item a measure of local probabilistic propagation (experimentally derived; function of fluid and other system parameters),
\end{inlineenum}
the model predicts the expected size distribution of avalanches.
In this article, we investigate how droplet configuration affects the avalanche dynamics. We find the mean size of these avalanches to depend non-trivially on how droplets are packed together. Large variations in the avalanche dynamics are observed when droplet packing are different, even when the other system properties (number of droplets, fluid properties, channel geometry, etc.) are kept constant.
Bidisperse emulsions show less variation in the dynamics and they are surprisingly more stable than monodisperse emulsions.
To get a systems-level understanding of how a given droplet-configuration either facilitates or impedes the propagation of an avalanche, we employ a graph-theoretic analysis, where emulsions are expressed as graphs. We find that the properties of the underlying graph, namely the mean degree and the algebraic connectivity, are well correlated with the observed avalanche dynamics. We exploit this dependence to derive a data-based model that predicts the expected avalanche sizes from the properties of the graph.
\end{abstract}

\maketitle


\section{Introduction}
\begin{figure*}
    \centering
    \includegraphics[width=\linewidth]{ 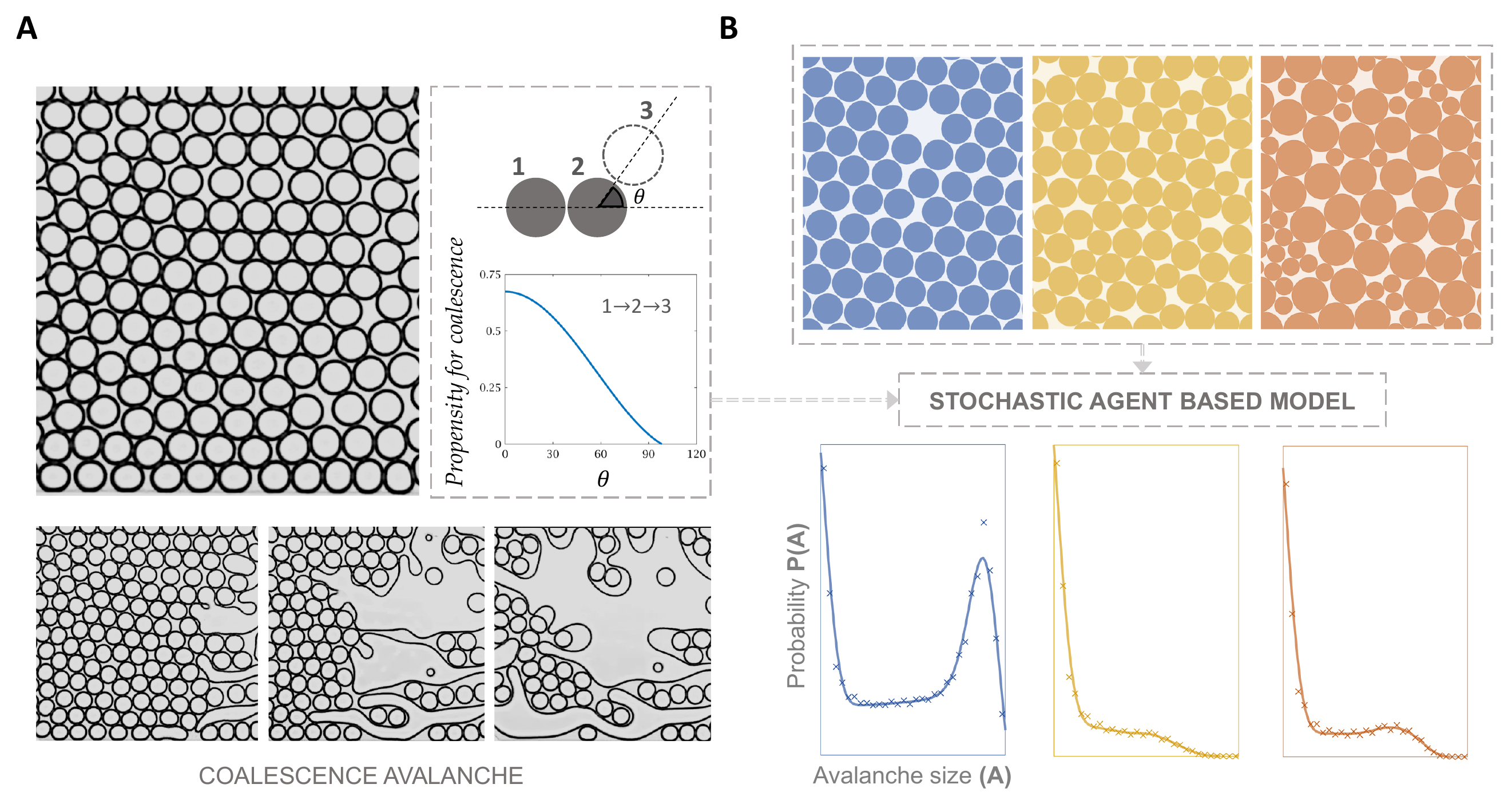}
    \caption{A - Snapshots from the experiments conducted by \cite{Bremond2011} (courtesy: Nicholas Bremond, ESPCI France). The large snapshot corresponds to the state before the onset of the avalanche. The smaller images show the time lapse snapshots of the propagating avalanche. The inset (in the box) illustrates the local propagation rule. When droplets 1 and 2 coalesce, a nearby droplet 3 coalesces with a probability based on its orientation $\theta$, as shown in the plot. B - Stochastic agent based model takes different droplet configurations as input, along with the local propagation rule in A, and predicts the probability of an avalanche via a Monte Carlo study for each of the configurations ($P(A)$) as a function of its size ($A$).}
    \label{fig:schematic-model}
\end{figure*}

Coalescence of a pair of droplets in a concentrated emulsion flowing through a 2D microchannel can trigger a cascade of similar events in their neighbourhood, giving rise to an avalanche that propagates through the emulsion \cite{Bremond2011, Gunes2010a} (See figure \ref{fig:schematic-model} A, for snapshots of the propagating coalescence avalanche). Since the coalescence process depends sensitively on various parameters like the film thickness of the liquid phase between the droplets, instantaneous velocities, etc. which vary dynamically across the emulsion, the avalanche is observed to propagate stochastically. Bremond et. al.~\cite{Bremond2011} even measured the probability associated with this local propagation as a function of the relative orientation of the droplets.
Using this measure in a stochastic agent based model, we simulate the propagation of coalescence avalanches in concentrated 2D emulsions. We find that avalanches either propagate autocatalytically to destabilise the entire emulsion, or prematurely stop cascading leaving it relatively stable~\cite{DannyRaj2017}. The avalanche dynamics depends on the size of the system, the aspect ratio of the packing, how the droplets are oriented with respect to each other locally~\cite{Bremond2011}, the number of avalanches triggered~\cite{DannyRaj2018}, fluid properties which in turn affect the overall propensity for coalescence propagation~\cite{DannyRaj2016, Baret2009}.

In our previous investigations of the phenomenon, we studied coalescence propagation only on closely packed assemblies of monodisperse droplets (of the same size). For these configurations, the total number of neighbours for each droplet were a constant except for those at the edge of the assembly. However, in real microfluidic applications, droplets self organise to form different arrangements, which could be randomly close-packed, with differences in the neighbour configuration between droplets even in the bulk of the assembly. Depending on the application, droplets may not always be monodisperse; and could be of different sizes. Polydisperse emulsions could significantly alter the neighbourhood configuration of droplets. An important question then arises: how sensitive is the avalanche dynamics to the underlying droplet configuration, when material composition and all other properties are kept the same?

If the propagation is indeed sensitive to how the droplets are packed, then in dense flowing conditions, like in droplet-based incubators \cite{Frenz2009, Dai2016b}, where droplets reorganise dynamically during the flow as they move through the channel (or engineered to do so), the stability of the emulsions, \textit{i.e.} the propensity to form large avalanches, becomes a function of time, making it hard to operate these devices stably.
Also, Bremond et al \cite{Bremond2011} hypothesised that polydisperse emulsions have a higher propensity to result in phase inversion inside a microchannel---where cascades of coalescence events could result in the inversion of the droplet and continuous phases locally in the emulsion. They showed that coalescence events leading to local phase inversion were limited in monodisperse emulsions due to the largely anisotropic propagation of the avalanche; for phases to invert, propagation should happen in small closed paths during which the continuous phase can be engulfed within the other coalescing phase. 

Therefore, to understand these systems in order to operate them stably or to control the overall propagation of coalescence avalanches, we need to investigate the role of droplet configuration on the emulsion stability.
To this end, we generate a variety of different droplet configurations and study their stability using a stochastic agent based model for coalescence propagation. 
Then, we bring graph theory to formally characterise the underlying droplet configuration---as a graph with \textit{nodes} representing droplets and \textit{edges} connecting droplets to their immediate neighbours---and investigate how the structural properties of the graph influences propagation. 
We then build a data-driven model that relate the topology of the droplet packing to emulsion stability.

\section{Modelling the stochastic coalescence avalanches.}

\subparagraph{Challenges with a first-principles approach.}
Coalescence of droplets is a multi-scale phenomenon~\cite{Chan2011, Janssen2011}. The continuous-phase film between two droplets has to drain and become thin enough to be unstable to perturbations which leads to its collapse that allows the droplet-interfaces to make contact. The drainage process is rather complicated: a range of different interface configurations are formed as the thin film drains~\cite{Vakarelski2010}. However, once these droplets touch, they form high curvature regions that lead to large surface tension forces that pull the droplets together. 

In the system we are interested in, droplets coalesce via a counter-intuitive mechanism: upon \textit{decompression}. After droplets that are sufficiently close to each other get pulled away, a low pressure region is formed that pulls the interfaces together initiating contact between the interfaces~\cite{Bremond2008}. 
Since these observations are reported in systems with large amounts of surfactant, there is reason to believe that coalescence is facilitated by the surfactant concentration gradients on the droplet interfaces.
To completely resolve these structures and capture the different stages as droplets coalesce, one requires very-fine time and space resolutions in their simulations.

Furthermore, to simulate a coalescence avalanche, one has to capture, 
\begin{inlineenum}
    \item self-organisation: the motion of droplets as they move through the microchannel, 
    \item nucleation events: coalescence between a pair of droplets which initiates a cascade of coalescence events,
    \item dynamic processes in coalescence: interactions between dynamically growing coalesced clusters formed due to multiple coalescence events.
\end{inlineenum}
These factors make any kind of first-principles approach to modelling coalescence avalanches computationally expensive, prohibiting the study at a system-level. 

\subparagraph{Need for a simple model.}
Our goal in this study is to understand how propagation of avalanches depend on the way droplets are packed together. 
Hence, what we need is a model that will take the droplet configuration as input and simulate the stochastic propagation of an avalanche.
The model should incorporate a measure of how coalescence events lead to newer events through the nearby droplets. 
For example, the probability associated with local propagation measured by Bremond and co-workers~\cite{Bremond2011} can be incorporated into such a model.
Also, since the avalanche propagates stochastically, it is important that we have a computationally simple model that allows us to generate independent realisations of the avalanche propagation (Monte Carlo study) to estimate the expected properties of the propagation phenomenon.

\subparagraph{Stochastic agent based model.} 
We model coalescence propagation as a stochastic branching process on a group of droplets packed together in a tight configuration~\cite{DannyRaj2016}; here a branch emerges when two droplets coalesce and the branch grows (or propagates) stochastically via neighbours that are in close proximity to the recently coalesced droplets. The process continues till all the newly formed branches either stop propagating or there are no more droplets to coalesce.
The droplets are assumed to be stationary during the entire propagation since, speeds associated with the propagation of the coalescence cascade is generally an order higher than the movement speeds.
Propagation is carried out on a randomly packed droplet configuration that is assembled using the algorithm in~\cite{Xu2005, Desmond2009}. We used the code provided by the authors of~\cite{Desmond2009} which can be found in~\cite{DesmondCodeRcp}, to produce dense mono- and bidisperse randomly packed droplet configurations (shown in figure~\ref{fig:schematic-model} B). Note that the stochastic agent based framework can be extended to account for the dynamic motion of the droplets and the coalesced clusters. However, in this article, we hold on to the simplifying assumption that droplets are static which aids our current interest: to understand how the avalanche dynamics depends on the configuration, of how droplets are packed.
For a detailed algorithm and implementation of the branching process the readers are referred to~\cite{DannyRaj2016}.

\subparagraph{Local propagation rule.} 
A pair of droplets is chosen randomly and allowed to coalesce. This initiates similar coalescence events in its neighbourhood with a probability $\alpha \times G(\theta)$; here, $\theta$ is a measure of the local orientation of the droplets participating in the propagation and $\alpha$ is a parameter that varies with fluid properties such as viscosity, surface tension, etc.
The form of $G(\theta)$ and the definition of $\theta$, are illustrated in figure~\ref{fig:schematic-model}~A. This measure was experimentally computed by ref~\cite{Bremond2011} after analysing over 2000 coalescence events in different parts of the 2D emulsion. The form resembles a cosine function, which is the component of the pulling force experienced by a new droplet due to coalescence of a pair of droplets~\cite{DannyRaj2016}.
$G(\theta)$ favours propagation along the orientation of the coalescing pair ($\theta = 0$), which gives rise to avalanches that propagate as fingers through the 2D emulsion.

\subparagraph{Fluid properties and critical transitions.}
The role of fluid characteristics on the propagation is implicitly captured using the parameter $\alpha$.
When $\alpha \simeq 1$, every new coalescence event has the means to initiate more such events through their neighbours resulting in a cascade of coalescence events (conditions same as the experiments of ref~\cite{Bremond2011}). When $\alpha$ is very small, coalescence events do not propagate and the emulsion is stable. Therefore, a critical $\alpha_c$ exists which marks this transition from system-size spanning avalanches to a stable regime. Similar qualitative transitions based on surfactant concentration were reported by Baret and co-workers~\cite{Baret2009}. 
We find that the structure of the observed $G(\theta)$, which favours finger-like propagation events leads to a system size dependence of the critical transition $\alpha_c = f(N)$ \cite{DannyRaj2017}.

\subparagraph{Monte Carlo study}
For every droplet configuration we perform a Monte-Carlo study ($\sim 10^5$ simulations) of the stochastic agent based model; every run generates an independent realisation of the stochastically propagating coalescence avalanche. From these independent runs, we compute the probability of occurrence of an avalanche $P(A)$, as a function of its size ($A$).

\section{Results and Discussion}
\begin{figure}
    \centering
    \includegraphics[width=\linewidth]{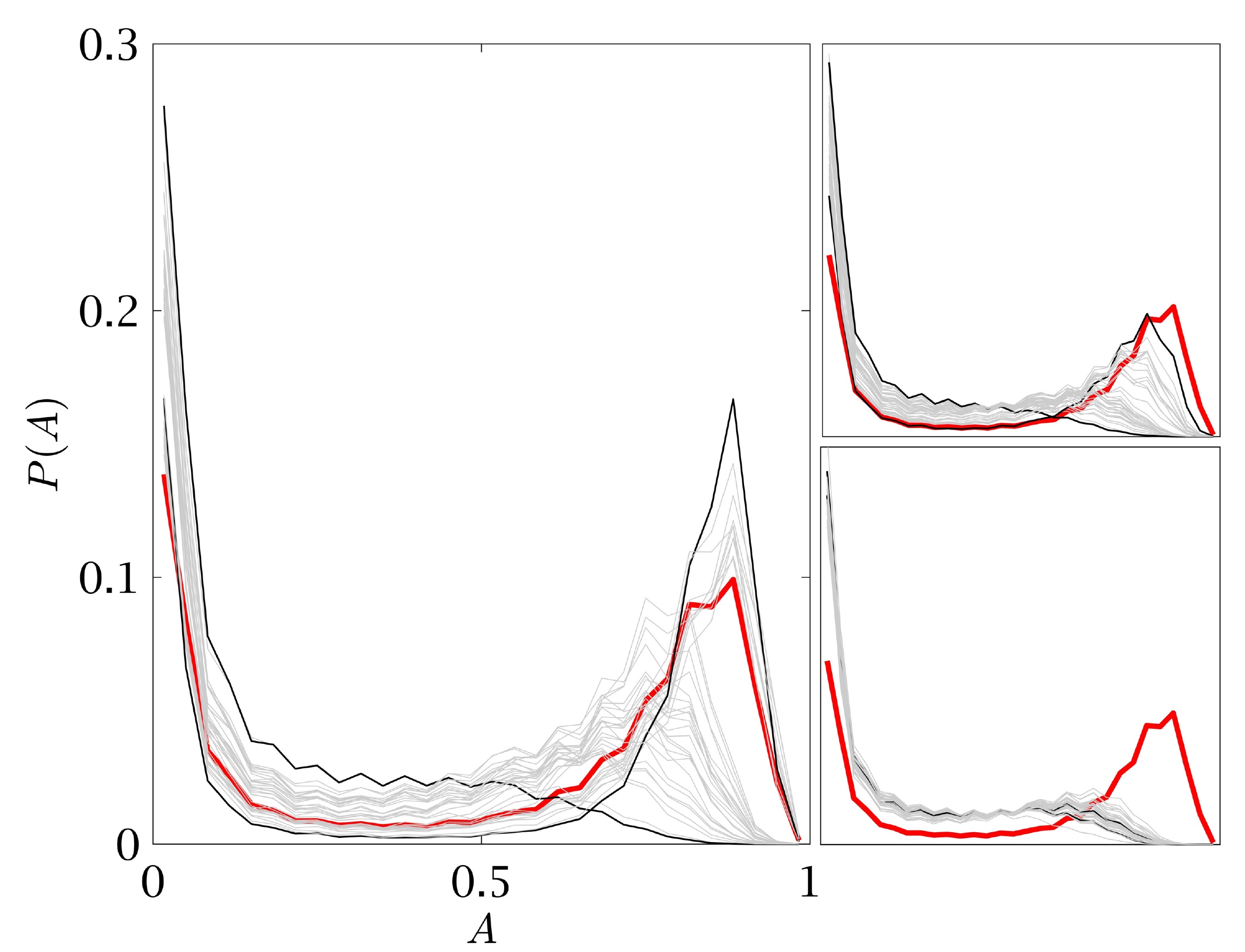}
    \caption{Probability of an avalanche $P(A)$ as a function of its size $A$ plotted for randomly packed monodisperse emulsion (main panel) and bidisperse emulsions (side panels). The top panel in the side corresponds to a small level of bidispersity ($30:70$ composition of droplets with size ratio $1.1:1$) and the bottom panel a large level of bidispersity ($50:50$ composition of droplets with size ratio $2:1$). These are same as the droplet configurations in figure \ref{fig:schematic-model} B. The thick red line is the $P(A)$ corresponding to the hexagonally close packed configuration (results from ref~\cite{DannyRaj2016}); the thin black lines correspond to the highest and lowest mean avalanches in a given category; the thin grey lines show the $P(A)$ for the rest. All the simulations here correspond to $\alpha = 1$.}
    \label{fig:AvalancheProbability_agentbasedmodel}
\end{figure}

\subparagraph{Stability of emulsions.}
The structure of $P(A)$, the probability of an avalanche of size $A$, sheds light on the nature of propagation and the resultant stability of the emulsion. In our previous investigations (\cite{DannyRaj2016, DannyRaj2018}), where the propagation was studied on a hexagonally close packed arrangement of droplets, we observed $P(A)$ to have a non-monotonic shape with a maximum value at very small values of $A$ and a second peak at a large value of $A$ (red curves in figure~\ref{fig:AvalancheProbability_agentbasedmodel}). The second peak indicates that a significant fraction of the avalanches propagate through the entire emulsion, destabilising it. One could call an emulsion `stable' if the second peak can be avoided. One way to do this would be to make $\alpha<\alpha_c$, which reduces the propensity associated with local propagation giving rise to coalescence events that do not propagate. 
This requires changing the fluids used, the surfactant concentration, etc.
Another way to reduce the second peak in $P(A)$, without having to change the fluid-system, is by changing the aspect ratio of the droplet configuration. Arranging the droplets in a slender configuration increases the chance of the propagating front to encounter the boundary more often, which reduces the chances associated with a system-level propagation.

\subparagraph{Propagation depends on the packing.}
When droplets of the same size are randomly packed, we observe significant variation in $P(A)$ across different droplet-configurations even when the parameters $\alpha$ and the aspect ratio of the droplet assembly are a constant (see thin lines in figure~\ref{fig:AvalancheProbability_agentbasedmodel}). The thick line (in red) corresponds to the monodisperse hexagonally close packed configuration (hcp) of droplets (same as the results in ref~\cite{DannyRaj2016}). 

When droplets are monodisperse and randomly packed, they exhibit the most variation about the hcp configuration, with few configurations even exhibiting a higher propagation than hcp.
However, when droplets are bidisperse, we find both the variation in $P(A)$ between different configurations and their mean propensity for propagation of large avalanches (second peak height) to decrease with increasing bidispersity (see side panels in figure~\ref{fig:AvalancheProbability_agentbasedmodel}). A configuration's level of bidispersity can be tuned by changing both the ratio of radii ($sr$) and the proportion of the two types of droplets ($nr$).

\subparagraph{Emulsions as graphs.}
When a pair of droplets coalesce, they lead to more such events in their neighbourhood. Hence, the number of neighbours available in their immediate vicinity and the angles they make with the recently coalesced pair determine the probabilities associated with propagation.
To understand the sustained propagation of an avalanche, one has to not just investigate the nearest neighbours, but also their neighbours, and so on. In other words, a system-level characterisation of the droplet configuration is essential to understand why a certain droplet-packing either favours or hinders the propagation of an avalanche.

\begin{figure}
    \centering
    \includegraphics[width=\linewidth]{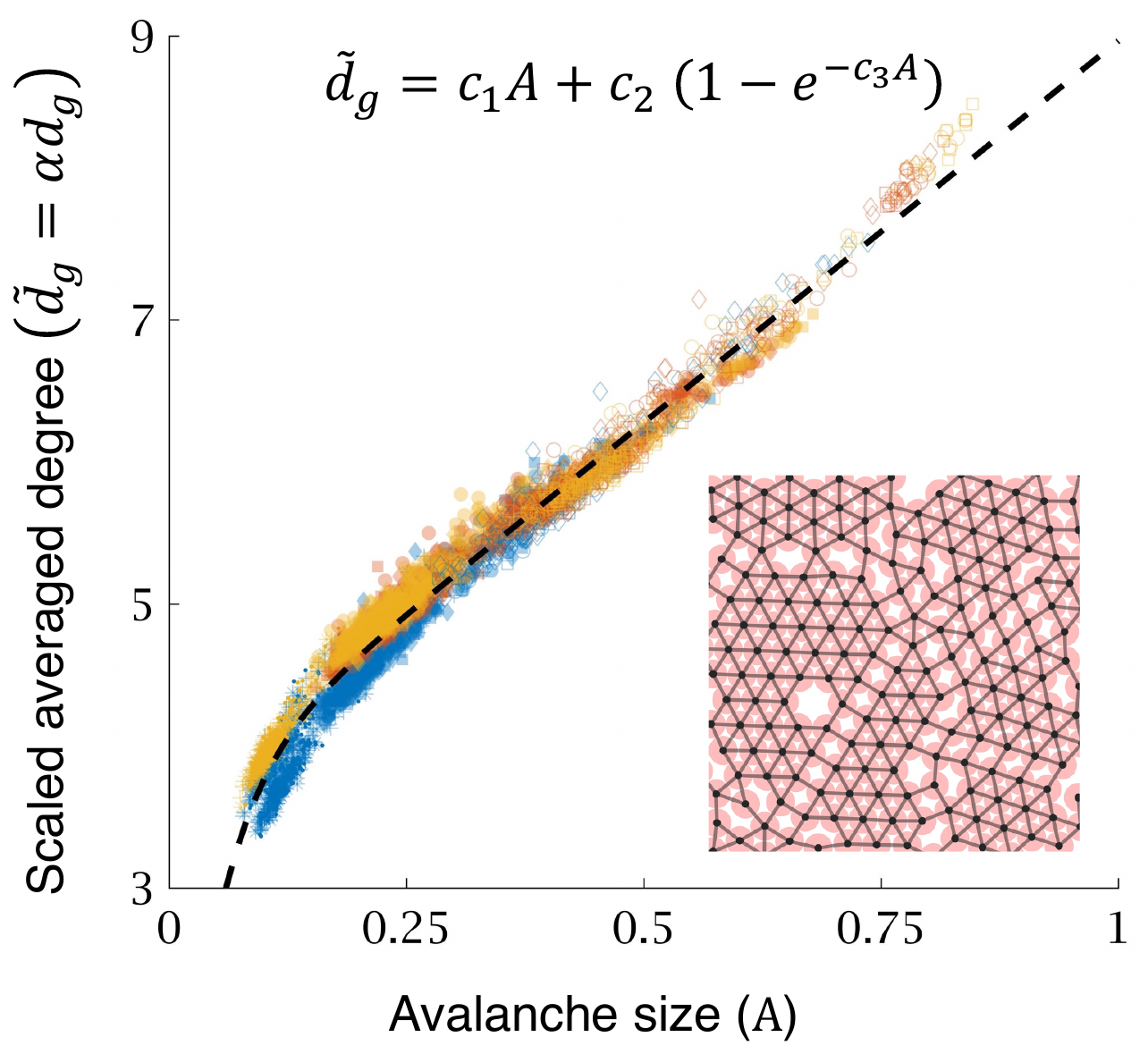}
    \caption{Average degree of a graph corresponding to a droplet configuration, scaled by the system parameter $\alpha$, as a function of the average avalanche size ($A$) it exhibits, plotted for different packing conditions: monodisperse, bidisperse with different size ratios ($sr$) and their compositions ($nr$), different total number of droplets ($N$) and different propensities for propagation ($\alpha$). The dotted line is a fit for a linear function with correction, $\Tilde{d}_g = c_1 A + c_2 (1-\exp^{-c_3 A})$, where $c_1 = 5.4$, $c_2 = 3.6$ and $c_3 = 22.9$. INSET: an example droplet configuration with an overlay of the corresponding graph. Parameters explored: $N \in \{144, 196, 225\}$, $\alpha \in \{0.9, 1, 1.1\}$, $sr \in (0, 3)$ and $nr \in (0, 0.5)$.}
    \label{fig:Emulsion_as_graphs}
\end{figure}

Tools from \textit{graph theory} offer a convenient formalism to analyse such system-level aspects of droplet configurations. A concentrated emulsion can be thought of as a \textit{graph} where the \textit{nodes} correspond to the droplets and an \textit{edge} connects every pair of nearby droplets through which coalescence can propagate (see inset of figure \ref{fig:Emulsion_as_graphs}). 
In this context, the question of interest takes the form: Can one understand the avalanche propagation dynamics from the topology of the underlying graph?

Coalescence avalanches can be interpreted as a cascade on the graph. If $x_i[t]$ is the probability that a node $i$ is `coalesced' at time $t$, then one can write the coalescence propagation equation as follows:

\begin{equation}
    x_i[t] = 1 - \prod_{j\in \mathcal{N}_i} \Big(1 - w_{ji} x_{j}[t-1]\Big)
    \label{eqn:markovchain}
\end{equation}
The RHS of Eq~\ref{eqn:markovchain} quantifies the probability associated with propagation from any of the possible neighbours of $i$.
Here, $w_{ji}$ is the probability that coalescence will propagate from $j$ to $i$; $\mathcal{N}_i$ refers to neighbourhood of $i$.
$w_{ji}$ is non-zero only when there is an edge connecting $j$ and $i$. This equation is similar in spirit to the time-evolution equation of a discrete-state, discrete-time Markov chain~\cite{ross2014probability}, where $x_i[t]$ is the probability of the Markov chain in state $i$. If $\mathbf{x} = [x_i]_{i=1}^N$ is the vector of all $x_i$, and $W = [w_{ij}]_{i, j = 1}^N$, then the propagation equation of the Markov chain can be written as $\mathbf x [t+1] = W \mathbf x [t]$. However, in the case of coalescence propagation, crucial properties such as the \emph{row-stochastic} nature of $W$, and the fact that $\sum_i x_i = 1$ do not hold---hence, we have a slightly more involved propagation equation given above.

A propagating coalescence avalanche could reach $i$ from $j$ through any of its neighbours $k$. Hence, we define $w_{ji}$ as,
\begin{equation}
    w_{ji}=
        \left\{\begin{matrix}
        \sum_{k\in\mathcal{N}_j} p(k,j,i) & A_{ji} = 1\\ 
        0 & otherwise
        \end{matrix}\right.
        \label{eqn:wij_defn}
\end{equation}
$p(k,j,i)$ is the probability that a coalescence event between droplets $k$ and $j$ results in the coalescence of droplet $i$. This is identical to the local probability rule, derived from observations of Bremond et al~\cite{Bremond2011} which is used in the stochastic agent based model (see figure \ref{fig:schematic-model} A). Since, the probability for propagation from $j$ to $i$ depends on the neighbours of $j$, we expect $w_{ji} \neq w_{ij}$.

One can immediately see, from the model according to eq \ref{eqn:markovchain}, that the propagation depends on how the droplets are packed. This is summarised in $W$, which can be thought of as a \emph{weighted adjacency matrix} of the underlying graph. The steady state distribution of an avalanche $\mathbf{x}^s$, \textit{i.e.} when $\mathbf{x}_i[t] = \mathbf{x}_i[t-1]=\mathbf{x}^s$, is an implicit function of the adjacency matrix $W$ alone. Hence, understanding the properties of the $W$ would indeed aid in understanding the role of droplet configuration on the avalanche propagation.

\subparagraph{Mean degree explains observed avalanches.}
A simple way to characterise a droplet configuration is via the \emph{degree distribution} of the underlying graph \cite{NetworksNeuman2010}. The degree of a node is equal to the number of nearby neighbours of the corresponding droplet, weighted by the probability that coalescence will propagate from that node. When computed for a weighted graph $w$, this corresponds to the net number of neighbours through which coalescence can effectively propagate. 
Randomly packing a small number of droplets could lead to large variations in the degree distributions between individual configurations that can in turn affect the propagation of avalanches, which is the reason for the variation in the $P(A)$ curves observed in figure \ref{fig:AvalancheProbability_agentbasedmodel} when droplets are randomly packed.

It is intuitive to expect the configurations with lower avalanche sizes to have smaller mean degree $d_g$ (averaged over all the droplets in a given configuration). We observe this to be generally true across configurations, even when they are bidisperse. 
We observe the mean degree, when scaled by the system parameter $\alpha$, ($\Tilde{d}_g = \alpha \times d_g$), to increase monotonically with the mean avalanche size $A$. We find all the data from our simulations---different configurations for both mono and bidisperse cases, different system sizes, and local propagation propensities $\alpha$---collapse on to a master curve, suggesting a universal relationship between the structure of the graph (defined based on the average degree of a configuration) and the expected avalanche size (averaged over many independent realisations) (see figure \ref{fig:Emulsion_as_graphs}). We build a relationship between $\Tilde{d}_g$ and $A$, from data.
\begin{equation}
    \Tilde{d}_g = c_1 A + c_2 \big(1 - e^{-c_3 A} \big).
    \label{eqn:data-driven-model}
\end{equation}
We regress a linear model with a correction term (for small $A$) to account for the phase transition (from autocatalytic to non-autocatalytic propagation) that corresponds with the small avalanche sizes. We find this relationship to hold true for different values of system size $N$, different propensities for propagation $\alpha$ and varying levels of bidispersity defined by the parameters $sr$ and $nr$---yielding a universal relationship.

\begin{figure}
    \centering
    \includegraphics[width=\linewidth]{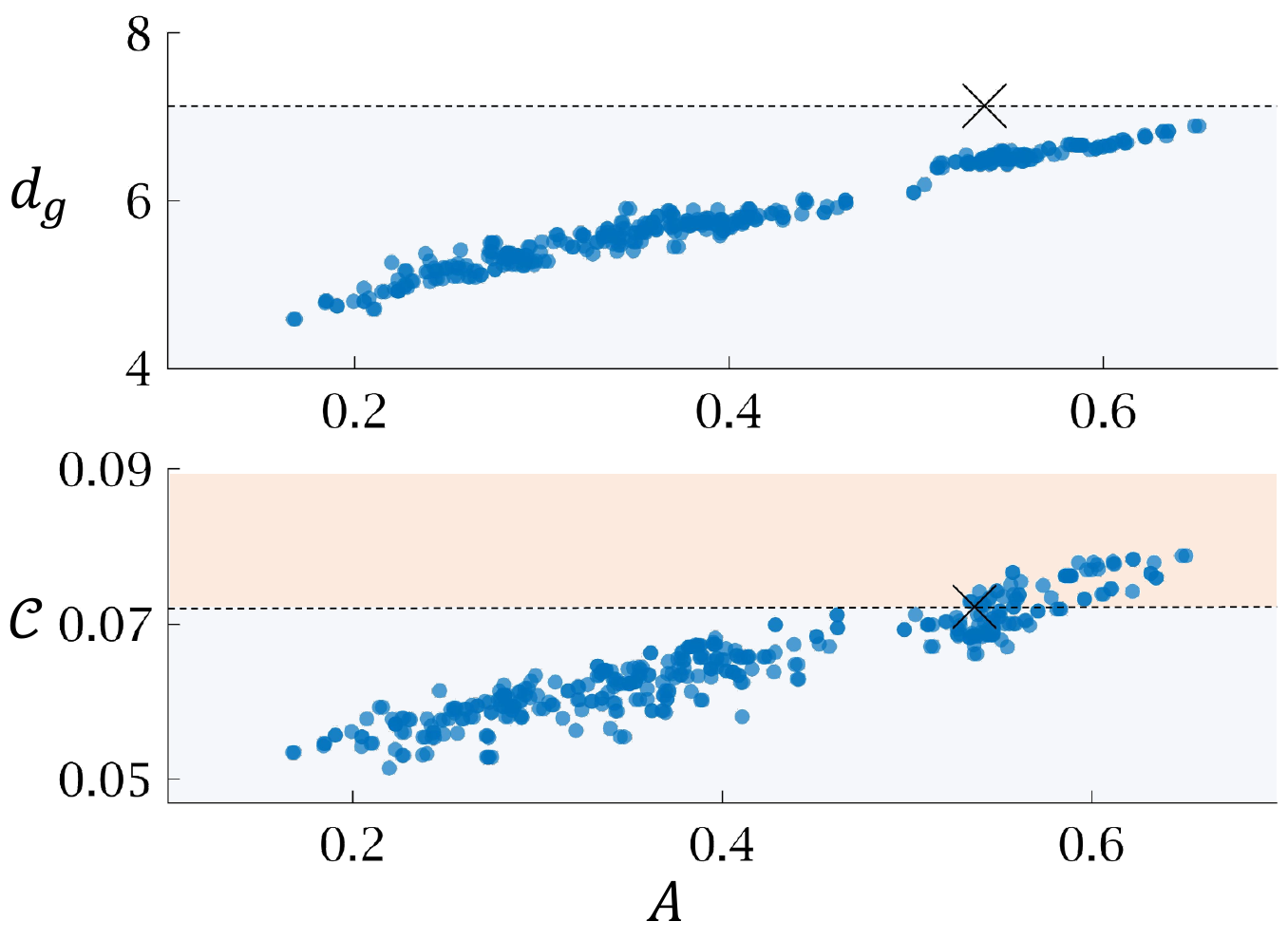}
    \caption{Average degree $d_g$ (top panel) and algebraic connectivity $\mathcal{C}$ (bottom panel) are plotted against the observed mean avalanche sizes $A$ for monodisperse, randomly packed droplet configurations along with that corresponding to hcp configuration (marked as \textbf{X}) for $\alpha = 1$. Top panel -- all the points lie in the region shaded which corresponds to the points that have a lower value of $d_g$ than hcp; bottom panel -- two regions are shaded: configurations with higher algebraic connectivity than hcp are in the top region exhibiting larger mean avalanches and \textit{vice versa}.}
    \label{fig:conductance_monodisperse}
\end{figure}

\subparagraph{Exceptions to the rule.}
While the data-driven model explains the dependence of the mean avalanche size to the packing characteristic $\Tilde{d_g}$ accurately for most cases, it does not explain why some of the droplet configurations corresponding to the monodisperse randomly packed emulsions facilitate avalanches larger than that by the hexagonally packed configuration (hcp) (figure \ref{fig:AvalancheProbability_agentbasedmodel} main panel and \ref{fig:conductance_monodisperse} top panel). This is puzzling, since a hcp configuration has the highest possible $d_g$ for a system of a given size.
While $d_g$ is small for these configurations with larger $A$, we find them to have different degree-distribution; the nodes with $6$ neighbours are not as high as hcp, however, they have more number of nodes with $5$ and $4$ neighbours. It is reasonable to expect that this difference in distribution is the cause for the better `flow' that promotes a growing avalanche.

To understand how a droplet configuration as a whole affects the propagation of the avalanche, we look at the \textit{conductance} of the graph, which is a measure that can be used to quantify how well-connected a graph is\cite{spielman2007spectral}. For instance, if a graph has bottlenecks, where there are only a small number of edges across the bottleneck, the conductance will be low. In the context of coalescence avalanches, presence of bottlenecks would decrease the propensity of the system to exhibit large avalanches.
Unfortunately, graph conductance as it is formally defined is algorithmically difficult to estimate~\cite{conductance}.
Hence, we use \emph{algebraic connectivity} $\mathcal{C}$, i.e. the second smallest eigenvalue of the Laplacian matrix of the graph, which is a proxy for the conductance \cite{spielman2007spectral, spielman2012spectral}. 
Conductance also has interpretations in terms of the diffusion time for diffusion processes on graphs, \textit{i.e} random walks. Since coalescence propagation is a stochastic process, we believe that conductance of a graph (measured as algebraic connectivity) would yield a single measure for the entire graph that can explain the large mean avalanche sizes observed.

\subparagraph{Algebraic connectivity explains the exceptions.}
In figure~\ref{fig:conductance_monodisperse}, we plot both the mean degree $d_g$ (in the top panel) and the algebraic connectivity $\mathcal{C}$ (in the bottom panel) for monodisperse randomly packed configurations as a function of $A$, the mean avalanche size.
We find that algebraic connectivity successfully relates the topology of the graph to the propensity for overall propagation (or flow) of the avalanches: $\mathcal{C}$ is found to increase linearly with $A$, similar to $d_g$. 
Further, algebraic connectivity also explains why some randomly packed configurations facilitate larger avalanches than hcp.
In figure~\ref{fig:conductance_monodisperse}, hcp configuration (marked as `X') has a higher $d_g$ in comparison to all the randomly packed configurations (top panel). However, the configurations that exhibit higher avalanche sizes than a hcp are also found to have greater algebraic connectivity values (bottom panel).


To understand how the algebraic connectivity, as a measure, works in successfully estimating a slightly lower flow in hcp, we examine its graph connectivity in a little more detail.
In the bulk of a hcp configuration, all droplets have the maximum possible number of neighbours (\textit{i.e.} 6). Hence, there are no `bottlenecks' in the bulk that could reduce the overall flow in a hcp.
However, the droplets near the boundary of the assembly have relatively low degrees (usually 4 or lower), while those for the randomly packed configurations have a wider degree distribution with a considerable number of droplets with a degree of 5.


Now, consider those avalanches that are triggered in the bulk, which propagate towards the boundary. These can become system-spanning only if the boundary is well-connected to the bulk.
Therefore, these avalanches have a higher chance of propagating further, in randomly packed configurations which have a net higher mean degree near the boundary than a hcp---a potential reason for the observed higher avalanche probabilities than hcp. Algebraic connectivity, being able to characterise the overall flow in the graph, is able to capture this phenomenon well.


Algebraic connectivity $\mathcal{C}$ could have been used as a regressor in the data-driven model (Eq \ref{eqn:data-driven-model}) instead of the scaled mean degree $\Tilde{d}_g$. However in our investigations, we find that using $\Tilde{d}_g$ achieves a more accurate fit to observed average avalanche size $A$ than using $\mathcal{C}$; compare the spread associated with the two metrics for a fixed $A$ in figure \ref{fig:conductance_monodisperse}.
While $\mathcal{C}$ explains the exceptions well, it is always affected by the existence of regions on the graph that contribute to smaller `flow'---since, $\mathcal{C}$ is a global measure. However, $\Tilde{d}_g$ being an average local measure of the graph topology, is not affected by these regions and serves as a better regressor in Eq~\ref{eqn:data-driven-model}.
Including both the metrics as regressors does not change the accuracy in explaining the data appreciably.

\section{Conclusion}
In this article, we show how the propensity for coalescence avalanches to propagate in a concentrated emulsion is directly tied to how the droplets are packed. By expressing the droplet configurations as graphs, we are able to employ graph theory formalism to study how the topology of droplet packing can be used to predict the size of coalescence avalanches that these configurations can facilitate.
We show the existence of a universal relationship between the scaled mean degree of the graph and the mean avalanche size, which allows us to build a data-driven model that can be used for predicting the avalanche size for a given droplet configuration. 
Since, the input to the data-driven model is simply how the droplets are packed, this model has the potential to be deployed online, in a real droplet microfluidic setup, to study the stability of packing as droplets self-organise in a 2D microfluidic channel. 
We find that there are exceptions to the data-driven model which is based on the mean degree of a graph---which is only an average local property of the graph. These are explained by algebraic connectivity, a proxy for graph-conductance, which that takes into account the system-level characteristics of a graph that are linked to a successful propagation of avalanches.

Also, since, we relate the topology of the droplet configuration via metrics that characterises its structure to the propagation dynamics, we expect our approach to be agnostic to the specific choice of the coalescence propagation model used in the study. Though specific details of the hydrodynamics are not considered explicitly, 
\begin{inlineenum}
    \item our approach takes into account the local propagation propensity (estimated from experiments) and extended to systems with different fluid properties and 
    \item the graph essentially represents the droplet-interaction network where every edge connects every pair of droplets that are within their sphere of hydrodynamic influence.
\end{inlineenum}
Hence, even the propagation dynamics predicted by a more detailed model---that considers the motion of droplets, thin-film thickness distribution and its effect on triggering a coalescence event, etc.---would still depend on how droplets are packed together and how this configuration gives rise to the propagation of the coalescence avalanches; which is captured by a graphical representation of the droplet ensembles.


\section*{Acknowledgments}
This project was funded by the DST INSPIRE faculty award, grant number: DST/INSPIRE/04/2017/002985. DRM acknowledges the technical assistance from Bhavya Balu and the discussions with Raghunathan Rengaswamy at IIT Madras, in the early stages of the project.

\bibliography{references.bib, references_2.bib}

\begin{thebibliography}{21}%
\makeatletter
\providecommand \@ifxundefined [1]{%
 \@ifx{#1\undefined}
}%
\providecommand \@ifnum [1]{%
 \ifnum #1\expandafter \@firstoftwo
 \else \expandafter \@secondoftwo
 \fi
}%
\providecommand \@ifx [1]{%
 \ifx #1\expandafter \@firstoftwo
 \else \expandafter \@secondoftwo
 \fi
}%
\providecommand \natexlab [1]{#1}%
\providecommand \enquote  [1]{``#1''}%
\providecommand \bibnamefont  [1]{#1}%
\providecommand \bibfnamefont [1]{#1}%
\providecommand \citenamefont [1]{#1}%
\providecommand \href@noop [0]{\@secondoftwo}%
\providecommand \href [0]{\begingroup \@sanitize@url \@href}%
\providecommand \@href[1]{\@@startlink{#1}\@@href}%
\providecommand \@@href[1]{\endgroup#1\@@endlink}%
\providecommand \@sanitize@url [0]{\catcode `\\12\catcode `\$12\catcode
  `\&12\catcode `\#12\catcode `\^12\catcode `\_12\catcode `\%12\relax}%
\providecommand \@@startlink[1]{}%
\providecommand \@@endlink[0]{}%
\providecommand \url  [0]{\begingroup\@sanitize@url \@url }%
\providecommand \@url [1]{\endgroup\@href {#1}{\urlprefix }}%
\providecommand \urlprefix  [0]{URL }%
\providecommand \Eprint [0]{\href }%
\providecommand \doibase [0]{https://doi.org/}%
\providecommand \selectlanguage [0]{\@gobble}%
\providecommand \bibinfo  [0]{\@secondoftwo}%
\providecommand \bibfield  [0]{\@secondoftwo}%
\providecommand \translation [1]{[#1]}%
\providecommand \BibitemOpen [0]{}%
\providecommand \bibitemStop [0]{}%
\providecommand \bibitemNoStop [0]{.\EOS\space}%
\providecommand \EOS [0]{\spacefactor3000\relax}%
\providecommand \BibitemShut  [1]{\csname bibitem#1\endcsname}%
\let\auto@bib@innerbib\@empty
\bibitem [{\citenamefont {Bremond}\ \emph {et~al.}(2011)\citenamefont
  {Bremond}, \citenamefont {Dom{\'{e}}jean},\ and\ \citenamefont
  {Bibette}}]{Bremond2011}%
  \BibitemOpen
  \bibfield  {author} {\bibinfo {author} {\bibfnamefont {N.}~\bibnamefont
  {Bremond}}, \bibinfo {author} {\bibfnamefont {H.}~\bibnamefont
  {Dom{\'{e}}jean}},\ and\ \bibinfo {author} {\bibfnamefont {J.}~\bibnamefont
  {Bibette}},\ }\bibfield  {title} {\bibinfo {title} {{Propagation of Drop
  Coalescence in a Two-Dimensional Emulsion: A Route towards Phase
  Inversion}},\ }\href {https://doi.org/10.1103/PhysRevLett.106.214502}
  {\bibfield  {journal} {\bibinfo  {journal} {Physical Review Letters}\
  }\textbf {\bibinfo {volume} {106}},\ \bibinfo {pages} {214502} (\bibinfo
  {year} {2011})}\BibitemShut {NoStop}%
\bibitem [{\citenamefont {Gunes}\ \emph {et~al.}(2010)\citenamefont {Gunes},
  \citenamefont {Clain}, \citenamefont {Breton}, \citenamefont {Mayor},\ and\
  \citenamefont {Burbidge}}]{Gunes2010a}%
  \BibitemOpen
  \bibfield  {author} {\bibinfo {author} {\bibfnamefont {D.~Z.}\ \bibnamefont
  {Gunes}}, \bibinfo {author} {\bibfnamefont {X.}~\bibnamefont {Clain}},
  \bibinfo {author} {\bibfnamefont {O.}~\bibnamefont {Breton}}, \bibinfo
  {author} {\bibfnamefont {G.}~\bibnamefont {Mayor}},\ and\ \bibinfo {author}
  {\bibfnamefont {A.~S.}\ \bibnamefont {Burbidge}},\ }\bibfield  {title}
  {\bibinfo {title} {{Avalanches of coalescence events and local extensional
  flows--stabilisation or destabilisation due to surfactant.}},\ }\href
  {https://doi.org/10.1016/j.jcis.2009.11.035} {\bibfield  {journal} {\bibinfo
  {journal} {Journal of colloid and interface science}\ }\textbf {\bibinfo
  {volume} {343}},\ \bibinfo {pages} {79} (\bibinfo {year} {2010})}\BibitemShut
  {NoStop}%
\bibitem [{\citenamefont {Danny~Raj}\ and\ \citenamefont
  {Rengaswamy}(2017)}]{DannyRaj2017}%
  \BibitemOpen
  \bibfield  {author} {\bibinfo {author} {\bibfnamefont {M.}~\bibnamefont
  {Danny~Raj}}\ and\ \bibinfo {author} {\bibfnamefont {R.}~\bibnamefont
  {Rengaswamy}},\ }\bibfield  {title} {\bibinfo {title} {{Averaged model for
  probabilistic coalescence avalanches in two-dimensional emulsions: Insights
  into uncertainty propagation}},\ }\href
  {https://doi.org/10.1103/PhysRevE.95.032608} {\bibfield  {journal} {\bibinfo
  {journal} {Physical Review E}\ }\textbf {\bibinfo {volume} {95}},\ \bibinfo
  {pages} {032608} (\bibinfo {year} {2017})}\BibitemShut {NoStop}%
\bibitem [{\citenamefont {Danny~Raj}\ and\ \citenamefont
  {Rengaswamy}(2018)}]{DannyRaj2018}%
  \BibitemOpen
  \bibfield  {author} {\bibinfo {author} {\bibfnamefont {M.}~\bibnamefont
  {Danny~Raj}}\ and\ \bibinfo {author} {\bibfnamefont {R.}~\bibnamefont
  {Rengaswamy}},\ }\bibfield  {title} {\bibinfo {title} {{Interacting
  coalescence avalanches in a 2D droplet assembly}},\ }\href
  {https://doi.org/10.1002/aic.16465} {\bibfield  {journal} {\bibinfo
  {journal} {AIChE Journal}\ }\textbf {\bibinfo {volume} {00}},\ \bibinfo
  {pages} {aic.16465} (\bibinfo {year} {2018})}\BibitemShut {NoStop}%
\bibitem [{\citenamefont {Danny~Raj}\ and\ \citenamefont
  {Rengaswamy}(2016)}]{DannyRaj2016}%
  \BibitemOpen
  \bibfield  {author} {\bibinfo {author} {\bibfnamefont {M.}~\bibnamefont
  {Danny~Raj}}\ and\ \bibinfo {author} {\bibfnamefont {R.}~\bibnamefont
  {Rengaswamy}},\ }\bibfield  {title} {\bibinfo {title} {{Coalescence of drops
  in a 2D microchannel: critical transitions to autocatalytic behaviour}},\
  }\href {https://doi.org/10.1039/C5SM01915J} {\bibfield  {journal} {\bibinfo
  {journal} {Soft Matter}\ }\textbf {\bibinfo {volume} {12}},\ \bibinfo {pages}
  {115} (\bibinfo {year} {2016})}\BibitemShut {NoStop}%
\bibitem [{\citenamefont {Baret}\ \emph {et~al.}(2009)\citenamefont {Baret},
  \citenamefont {Kleinschmidt}, \citenamefont {Harrak},\ and\ \citenamefont
  {Griffiths}}]{Baret2009}%
  \BibitemOpen
  \bibfield  {author} {\bibinfo {author} {\bibfnamefont {J.~C.}\ \bibnamefont
  {Baret}}, \bibinfo {author} {\bibfnamefont {F.}~\bibnamefont {Kleinschmidt}},
  \bibinfo {author} {\bibfnamefont {A.~E.}\ \bibnamefont {Harrak}},\ and\
  \bibinfo {author} {\bibfnamefont {A.~D.}\ \bibnamefont {Griffiths}},\
  }\bibfield  {title} {\bibinfo {title} {{Kinetic aspects of emulsion
  stabilization by surfactants: A microfluidic analysis}},\ }\href
  {https://doi.org/10.1021/la9000472} {\bibfield  {journal} {\bibinfo
  {journal} {Langmuir}\ }\textbf {\bibinfo {volume} {25}},\ \bibinfo {pages}
  {6088} (\bibinfo {year} {2009})}\BibitemShut {NoStop}%
\bibitem [{\citenamefont {Frenz}\ \emph {et~al.}(2009)\citenamefont {Frenz},
  \citenamefont {Blank}, \citenamefont {Brouzes},\ and\ \citenamefont
  {Griffiths}}]{Frenz2009}%
  \BibitemOpen
  \bibfield  {author} {\bibinfo {author} {\bibfnamefont {L.}~\bibnamefont
  {Frenz}}, \bibinfo {author} {\bibfnamefont {K.}~\bibnamefont {Blank}},
  \bibinfo {author} {\bibfnamefont {E.}~\bibnamefont {Brouzes}},\ and\ \bibinfo
  {author} {\bibfnamefont {A.~D.}\ \bibnamefont {Griffiths}},\ }\bibfield
  {title} {\bibinfo {title} {{Reliable microfluidic on-chip incubation of
  droplets in delay-lines.}},\ }\href {https://doi.org/10.1039/b816049j}
  {\bibfield  {journal} {\bibinfo  {journal} {Lab on a chip}\ }\textbf
  {\bibinfo {volume} {9}},\ \bibinfo {pages} {1344} (\bibinfo {year}
  {2009})}\BibitemShut {NoStop}%
\bibitem [{\citenamefont {Dai}\ \emph {et~al.}(2016)\citenamefont {Dai},
  \citenamefont {Kim}, \citenamefont {Guzman}, \citenamefont {Shim},\ and\
  \citenamefont {Han}}]{Dai2016b}%
  \BibitemOpen
  \bibfield  {author} {\bibinfo {author} {\bibfnamefont {J.}~\bibnamefont
  {Dai}}, \bibinfo {author} {\bibfnamefont {H.~S.}\ \bibnamefont {Kim}},
  \bibinfo {author} {\bibfnamefont {A.~R.}\ \bibnamefont {Guzman}}, \bibinfo
  {author} {\bibfnamefont {W.-B.}\ \bibnamefont {Shim}},\ and\ \bibinfo
  {author} {\bibfnamefont {A.}~\bibnamefont {Han}},\ }\bibfield  {title}
  {\bibinfo {title} {{A large-scale on-chip droplet incubation chamber enables
  equal microbial culture time}},\ }\href {https://doi.org/10.1039/C5RA26505C}
  {\bibfield  {journal} {\bibinfo  {journal} {RSC Advances}\ }\textbf {\bibinfo
  {volume} {6}},\ \bibinfo {pages} {20516} (\bibinfo {year}
  {2016})}\BibitemShut {NoStop}%
\bibitem [{\citenamefont {Chan}\ \emph {et~al.}(2011)\citenamefont {Chan},
  \citenamefont {Klaseboer},\ and\ \citenamefont {Manica}}]{Chan2011}%
  \BibitemOpen
  \bibfield  {author} {\bibinfo {author} {\bibfnamefont {D.~Y.~C.}\
  \bibnamefont {Chan}}, \bibinfo {author} {\bibfnamefont {E.}~\bibnamefont
  {Klaseboer}},\ and\ \bibinfo {author} {\bibfnamefont {R.}~\bibnamefont
  {Manica}},\ }\bibfield  {title} {\bibinfo {title} {{Film drainage and
  coalescence between deformable drops and bubbles}},\ }\href
  {https://doi.org/10.1039/c0sm00812e} {\bibfield  {journal} {\bibinfo
  {journal} {Soft Matter}\ }\textbf {\bibinfo {volume} {7}},\ \bibinfo {pages}
  {2235} (\bibinfo {year} {2011})}\BibitemShut {NoStop}%
\bibitem [{\citenamefont {Janssen}\ and\ \citenamefont
  {Anderson}(2011)}]{Janssen2011}%
  \BibitemOpen
  \bibfield  {author} {\bibinfo {author} {\bibfnamefont {P.~J.~a.}\
  \bibnamefont {Janssen}}\ and\ \bibinfo {author} {\bibfnamefont {P.~D.}\
  \bibnamefont {Anderson}},\ }\bibfield  {title} {\bibinfo {title} {{Modeling
  Film Drainage and Coalescence of Drops in a Viscous Fluid}},\ }\href
  {https://doi.org/10.1002/mame.201000375} {\bibfield  {journal} {\bibinfo
  {journal} {Macromolecular Materials and Engineering}\ }\textbf {\bibinfo
  {volume} {296}},\ \bibinfo {pages} {238} (\bibinfo {year}
  {2011})}\BibitemShut {NoStop}%
\bibitem [{\citenamefont {Vakarelski}\ \emph {et~al.}(2010)\citenamefont
  {Vakarelski}, \citenamefont {Manica}, \citenamefont {Tang}, \citenamefont
  {O'Shea}, \citenamefont {Stevens}, \citenamefont {Grieser}, \citenamefont
  {Dagastine},\ and\ \citenamefont {Chan}}]{Vakarelski2010}%
  \BibitemOpen
  \bibfield  {author} {\bibinfo {author} {\bibfnamefont {I.~U.}\ \bibnamefont
  {Vakarelski}}, \bibinfo {author} {\bibfnamefont {R.}~\bibnamefont {Manica}},
  \bibinfo {author} {\bibfnamefont {X.}~\bibnamefont {Tang}}, \bibinfo {author}
  {\bibfnamefont {S.~J.}\ \bibnamefont {O'Shea}}, \bibinfo {author}
  {\bibfnamefont {G.~W.}\ \bibnamefont {Stevens}}, \bibinfo {author}
  {\bibfnamefont {F.}~\bibnamefont {Grieser}}, \bibinfo {author} {\bibfnamefont
  {R.~R.}\ \bibnamefont {Dagastine}},\ and\ \bibinfo {author} {\bibfnamefont
  {D.~Y.~C.}\ \bibnamefont {Chan}},\ }\bibfield  {title} {\bibinfo {title}
  {{Dynamic interactions between microbubbles in water.}},\ }\href
  {https://doi.org/10.1073/pnas.1005937107} {\bibfield  {journal} {\bibinfo
  {journal} {Proceedings of the National Academy of Sciences of the United
  States of America}\ }\textbf {\bibinfo {volume} {107}},\ \bibinfo {pages}
  {11177} (\bibinfo {year} {2010})}\BibitemShut {NoStop}%
\bibitem [{\citenamefont {Bremond}\ \emph {et~al.}(2008)\citenamefont
  {Bremond}, \citenamefont {Thiam},\ and\ \citenamefont
  {Bibette}}]{Bremond2008}%
  \BibitemOpen
  \bibfield  {author} {\bibinfo {author} {\bibfnamefont {N.}~\bibnamefont
  {Bremond}}, \bibinfo {author} {\bibfnamefont {A.~R.}\ \bibnamefont {Thiam}},\
  and\ \bibinfo {author} {\bibfnamefont {J.}~\bibnamefont {Bibette}},\
  }\bibfield  {title} {\bibinfo {title} {{Decompressing Emulsion Droplets
  Favors Coalescence}},\ }\href
  {https://doi.org/10.1103/PhysRevLett.100.024501} {\bibfield  {journal}
  {\bibinfo  {journal} {Physical Review Letters}\ }\textbf {\bibinfo {volume}
  {100}},\ \bibinfo {pages} {024501} (\bibinfo {year} {2008})}\BibitemShut
  {NoStop}%
\bibitem [{\citenamefont {Xu}\ \emph {et~al.}(2005)\citenamefont {Xu},
  \citenamefont {Blawzdziewicz},\ and\ \citenamefont {O’Hern}}]{Xu2005}%
  \BibitemOpen
  \bibfield  {author} {\bibinfo {author} {\bibfnamefont {N.}~\bibnamefont
  {Xu}}, \bibinfo {author} {\bibfnamefont {J.}~\bibnamefont {Blawzdziewicz}},\
  and\ \bibinfo {author} {\bibfnamefont {C.}~\bibnamefont {O’Hern}},\
  }\bibfield  {title} {\bibinfo {title} {{Random close packing revisited: Ways
  to pack frictionless disks}},\ }\href
  {https://doi.org/10.1103/PhysRevE.71.061306} {\bibfield  {journal} {\bibinfo
  {journal} {Physical Review E}\ }\textbf {\bibinfo {volume} {71}},\ \bibinfo
  {pages} {061306} (\bibinfo {year} {2005})}\BibitemShut {NoStop}%
\bibitem [{\citenamefont {Desmond}\ and\ \citenamefont
  {Weeks}(2009{\natexlab{a}})}]{Desmond2009}%
  \BibitemOpen
  \bibfield  {author} {\bibinfo {author} {\bibfnamefont {K.~W.}\ \bibnamefont
  {Desmond}}\ and\ \bibinfo {author} {\bibfnamefont {E.~R.}\ \bibnamefont
  {Weeks}},\ }\bibfield  {title} {\bibinfo {title} {{Random close packing of
  disks and spheres in confined geometries}},\ }\href
  {https://doi.org/10.1103/PhysRevE.80.051305} {\bibfield  {journal} {\bibinfo
  {journal} {Physical Review E}\ }\textbf {\bibinfo {volume} {80}},\ \bibinfo
  {pages} {051305} (\bibinfo {year} {2009}{\natexlab{a}})}\BibitemShut
  {NoStop}%
\bibitem [{\citenamefont {Desmond}\ and\ \citenamefont
  {Weeks}(2009{\natexlab{b}})}]{DesmondCodeRcp}%
  \BibitemOpen
  \bibfield  {author} {\bibinfo {author} {\bibfnamefont {K.~W.}\ \bibnamefont
  {Desmond}}\ and\ \bibinfo {author} {\bibfnamefont {E.~R.}\ \bibnamefont
  {Weeks}},\ }\href
  {http://www.physics.emory.edu/faculty/weeks/ken/ConfinedRCP.html} {\bibinfo
  {title} {{Code: To generate randomly closed packed discs}}} (\bibinfo {year}
  {2009}{\natexlab{b}})\BibitemShut {NoStop}%
\bibitem [{\citenamefont {Ross}(2014)}]{ross2014probability}%
  \BibitemOpen
  \bibfield  {author} {\bibinfo {author} {\bibfnamefont {S.~M.}\ \bibnamefont
  {Ross}},\ }\href@noop {} {\emph {\bibinfo {title} {Introduction to
  Probability Models}}}\ (\bibinfo  {publisher} {Academic press},\ \bibinfo
  {year} {2014})\BibitemShut {NoStop}%
\bibitem [{\citenamefont {Newman}(2010)}]{NetworksNeuman2010}%
  \BibitemOpen
  \bibfield  {author} {\bibinfo {author} {\bibfnamefont {M.}~\bibnamefont
  {Newman}},\ }\href@noop {} {\emph {\bibinfo {title} {{Networks: An
  introduction}}}},\ \bibinfo {edition} {1st}\ ed.\ (\bibinfo  {publisher}
  {Oxford},\ \bibinfo {address} {New York, NY},\ \bibinfo {year}
  {2010})\BibitemShut {NoStop}%
\bibitem [{\citenamefont {Spielman}(2007)}]{spielman2007spectral}%
  \BibitemOpen
  \bibfield  {author} {\bibinfo {author} {\bibfnamefont {D.~A.}\ \bibnamefont
  {Spielman}},\ }\bibfield  {title} {\bibinfo {title} {Spectral graph theory
  and its applications},\ }in\ \href@noop {} {\emph {\bibinfo {booktitle} {48th
  Annual IEEE Symposium on Foundations of Computer Science (FOCS'07)}}}\
  (\bibinfo {organization} {IEEE},\ \bibinfo {year} {2007})\ pp.\ \bibinfo
  {pages} {29--38}\BibitemShut {NoStop}%
\bibitem [{con()}]{conductance}%
  \BibitemOpen
  \href@noop {} {}\bibinfo {note} {The conductance of a given cut of a graph is
  defined as the ratio of the number of edges across the cut to the number of
  edges within the smaller side of the cut. The conductance of the whole graph
  is the maximum value of conductance over all cuts. \cite{bollobas1998modern}
  Computing the conductance of a graph is an NP-hard problem.}\BibitemShut
  {Stop}%
\bibitem [{\citenamefont {Spielman}(2012)}]{spielman2012spectral}%
  \BibitemOpen
  \bibfield  {author} {\bibinfo {author} {\bibfnamefont {D.}~\bibnamefont
  {Spielman}},\ }\bibfield  {title} {\bibinfo {title} {Spectral graph theory},\
  }\href@noop {} {\bibfield  {journal} {\bibinfo  {journal} {Combinatorial
  scientific computing}\ }\textbf {\bibinfo {volume} {18}} (\bibinfo {year}
  {2012})}\BibitemShut {NoStop}%
\bibitem [{\citenamefont {Bollob{\'a}s}(1998)}]{bollobas1998modern}%
  \BibitemOpen
  \bibfield  {author} {\bibinfo {author} {\bibfnamefont {B.}~\bibnamefont
  {Bollob{\'a}s}},\ }\href@noop {} {\emph {\bibinfo {title} {Modern graph
  theory}}},\ Vol.\ \bibinfo {volume} {184}\ (\bibinfo  {publisher} {Springer
  Science \& Business Media},\ \bibinfo {year} {1998})\BibitemShut {NoStop}%
\end{thebibliography}%

\end{document}